\documentclass[aps,prl,showkeys,showpacs,preprint,groupedaddress]{revtex4}
\usepackage{graphicx}

\begin{document}

\title{Photons uncertainty solves Einstein-Podolsky-Rosen paradox}

\author{Daniele Tommasini}

\affiliation{Departamento de F\'\i sica Aplicada, \'Area de F\'\i
sica Te\'orica, Universidad de Vigo, 32004 Ourense, Spain}

\email[]{daniele@uvigo.es}

\date{\today}

\begin{abstract}
Einstein, Podolsky and Rosen (EPR) pointed out that the
quantum-mechanical description of ``physical reality" implied an
unphysical, instantaneous action between distant measurements. To
avoid such an action at a distance, EPR concluded that Quantum
Mechanics had to be incomplete. However, its extensions involving
additional ``hidden variables", allowing for the recovery of
determinism and locality, have been disproved experimentally
(Bell's theorem). In this talk, I present an opposite solution of
the paradox based on the greater indeterminism of the modern
Quantum Field Theory (QFT) description of Particle Physics, which
prevents the preparation of any state having a definite number of
particles. The resulting uncertainty in photons radiation has
interesting consequences in Quantum Information Theory (e.g.˜
cryptography and teleportation). Moreover, since it allows for
less elements of EPR physical reality than the old
non-relativistic Quantum Mechanics, QFT satisfies the EPR
condition of completeness without the need for hidden variables.
The residual physical reality doesn't ever violate locality, thus
the unique objective proof of ``quantum nonlocality" is removed in
an interpretation-independent way. On the other hand, the supposed
nonlocality of the EPR correlations turns out to be a problem of
the interpretation of the measurement process. If we do not rely
on hidden variables or new physics beyond QFT, the viable
interpretation is a minimal statistical one, which preserves
locality and Lorentz symmetry.
\end{abstract}

\pacs{03.70.+k; 03.65.Ta; 03.65.Bz; 11.10.-z; 11.15.-q; 11.30.Cp}

\keywords{EPR paradox; Quantum Field Theory; Quantum
Electrodynamics; Lorentz symmetry; Locality; Quantum Measurement;
Entanglement; Quantum Cryptography; Quantum Information}

\maketitle

In the ``orthodox" interpretation of Quantum Mechanics (QM), the
physical quantities are usually not defined until they are
measured; they have no ``physical reality", according to the
famous definition proposed in 1935 by Einstein, Podolsky and Rosen
(EPR) \cite{EPR}:

{\it ``If, without in any way disturbing a system we can predict
with certainty (i.e.˜ with probability equal to unity) the value
of a physical quantity, then there is an element of physical
reality corresponding to this physical quantity".} EPR explicitly
asked for a unit probability for this prediction, but we can also
replace their requirement with the weaker one: {\it with
probability $1-\epsilon$, where $\epsilon$ may be made arbitrarily
small} \cite{Ballentine70}.

However, EPR pointed out that even ordinary QM allows for some
elements of physical reality: if a system is prepared in an
eigenstate of a given observable, we can predict with certainty
that the result of the measurement of that observable will be the
corresponding eigenvalue: there is then an element of objective
physical reality corresponding to that observable.

In Classical Physics, all the physical quantities have a (possibly
unknown) definite value in a given system at a given time. In the
orthodox interpretation of QM, however, it is supposed that the
state vector completely describes the state of the considered
single system, and this does not allow for a certain prediction of
the results of the measurements of two noncommuting observables,
such as the position and the momentum. Given a state, there are
unavoidably some observables (heuristically, ``half" the set of
observables) that cannot have a reality, and their measurement in
an ensemble of copies of the system prepared in this state will
show a nonvanishing dispersion. Are these non-diagonalized
physical quantities really undefined on the single copy of the
system? Or is this uncertainty merely a consequence of an
inevitable lack of knowledge of some ``hidden variables" that
would allow for a complete description of the single system and
possibly for an underlying determinism? In other words, can QM be
completed to a kind of statistical mechanics?

The brilliant argument developed by EPR was aimed at resolving
this dilemma. First, they proposed their famous ``{\it condition
of completeness}":

``{\it every element of the physical reality must have a
counterpart in the physical theory}" \cite{EPR}.

Secondly, they invented a thought-experiment for which QM
predicted that two incompatible observables were given a
simultaneous reality. Hereafter, I will consider a variant due to
Bohm, the so-called ``EPR-Bohm" thought-experiment \cite{pureprp}.
Two charged spin $1/2$ particles (e.g.˜ an electron-positron
pair), A and B, are created in coincidence in a spin-singlet
state, described by the ``entangled" spin vector
\begin{equation}
\vert\psi\rangle=\frac{1}{\sqrt{2}}\left(\vert+\rangle_{A}\vert-\rangle_{B}
- \vert-\rangle_{A}\vert+\rangle_{B}\right), \label{entanglement}
\end{equation}
where e.g.˜ $\vert\pm\rangle_A$ are the usual eigenstates of the
spin component $S_z(A)$ of particle A, with eigenvalues
$\pm\hbar/2$ respectively. The two particles are then detected by
the detectors $O_A$ and $O_B$ in opposite directions. The
measurement of a given component $\vec S\cdot\vec a$ of the spin
of particle A (or of B) along a unit vector $\vec a$ can give the
values $\pm\hbar/2$, each with probability $1/2$. However, if
$\vec S\cdot\vec a$ is measured on A and found, say, equal to
$+\hbar/2$, then the value of the same spin component on B will be
known with certainty to be $-\hbar/2$ due to Eq.˜
(\ref{entanglement}).

EPR assumed that {\it the physical reality on B is independent of
what is done with A, which is spatially separated from the former}
\cite{EPR} (this assumption has been called {\it Einstein's
Locality}). Since a certain prediction for the considered spin
component on B was allowed {\it without in any way disturbing
particle B}, they concluded that  {\it there is an element of
physical reality for the spin component $\vec S\cdot\vec a$ of B}.
By repeating this argument for any component of the spin, they
deduced that {\it all the spin components ($S_x$, $S_y$, $S_z$)
must be given a simultaneous physical reality} on particle B. But
this contradicts ordinary QM as based on the assumption that the
wave function provides a complete description of the single
system, implying that only one component of the spin of a given
particle can have a sharp value.

In other words, as Einstein himself noticed in 1949
\cite{Einstein49}: ``The paradox forces us to relinquish one of
the following two assertions:"

1) the description by means of the wave function is {\it
complete},

2) the physical realities of spatially separated objects are
independent of each other.

The ``EPR paradox", defined as the incompatibility of statements
1) and 2), has also been called {\it ``EPR theorem"} (see e.g.˜
Refs.˜ \cite{Ballentine70,Peres,Laloe}).

Since ``Einstein's locality" assertion 2) was considered
unquestionable by EPR, they deduced that the wave function did not
provide a complete description of the state of a system. This was
a strong, objective argument for hidden variables, that would
allow for locality and possibly for determinism. In the resulting
extensions of QM, all the observables (such as $S_x$, $S_y$,
$S_z$) have a definite value in the single system that is under
consideration. The dispersion of the probability distributions
observed in the repetition of the experiment on an ensemble of
identically prepared systems is merely a ``statistical mechanics"
effect. The hidden determinism of the theory would explain the
fact that the measurement of a component of the spin of A
apparently has a deterministic effect on the distant measurement
of the same component of the spin of B: both results would
actually be the deterministic consequence of the common production
of the two particles.

However, Bell pointed out that local hidden variable theories
could not reproduce all the results of QM \cite{Bell64}. He
proposed a set of ``Bell's inequalities" for the spin correlations
in a realization of the EPR-Bohm experiment, that were violated by
QM and respected by any local deterministic hidden variable
theory. Since the actual experiments \cite{Aspect} confirmed the
predictions of QM, hidden variables allowing for local determinism
were ruled out. {\it This result will be called hereafter the
``original Bell's theorem".}

Therefore, it was deduced that QM was a complete theory, and the
EPR Theorem mentioned above led to the conclusion that it had to
be a ``nonlocal" theory. Hereafter, I will call this argument the
{\it ``EPR+Bell" proof of nonlocality,} since it is based on both
EPR and Bell's theorems.

Moreover, in the last several years there has been a proliferation
of {\it generalized ``Bell's Theorems",} claiming that the
observed violation of  Bell's inequalities was sufficient in
itself to prove the existence of an instantaneous influence
between distant measurements.

However, any kind of influence at a distance would be described as
an effect of the future on the past by suitable inertial observers
\cite{BaJa87}. In fact, the supposed ``quantum nonlocality" is the
main origin of the widespread belief that the Quantum Theory is
incompatible with Special Relativity \cite{measrel,Scarani},
although it is recognized that the EPR correlations do not allow
for superluminal signaling.

Here, I will point out that the EPR incompleteness argument and
the EPR theorem mentioned above can be removed in an
interpretation-independent way in the modern Quantum Field Theory
(QFT) description of Particle Physics \cite{pureprp}. The point is
that {\it it is impossible to prepare a state with a definite
number of particles, as assumed in the EPR argument, since
additional real particles are \underline{always} allowed to be
created. The predicted rate for the production of a given number
of additional particles of given kinds is a fixed, finite number,
that depends on the considered original system and cannot be made
arbitrarily small.} Which additional species can appear depends on
the available energy and on the conservation laws. Since massless
particles can have arbitrarily low energy, the possible presence
of real photons -which do not carry any conserved charge- should
{\it always} be taken into account in the theoretical treatment.
In Ref.˜ \cite{pureprp} I have given a very general proof of this
statement, valid for all kinds of EPR experiments, including those
using neutral particles such as photons, neutrinos, K or B mesons.
The result is that {\it an undetermined number of photons is
created in any experiment, in any step that involves an
interaction,} and in particular in the process that originates the
two (or more) particles involved in an EPR experiment
\cite{pureprp}. The resulting rate for the production of $n$
additional photons is typically of order $\alpha^n$ as compared to
that for the production of particles A and B alone. As I have
discussed in Ref.˜ \cite{pureprp}, this generality is by no means
accidental, but it is a consequence of the local gauge symmetry.

Although QFT is best formulated in terms of Green functions,
giving the amplitudes for scattering and decay processes, I will
summarize the result of Ref.˜ \cite{pureprp} by introducing an
effective state vector to statistically describe the system
arising from the given production process (to be more precise we
should use a density operator, since the state is not ``pure"). In
the case of our EPR-Bohm experiment, QFT predicts that the state
that is produced is
\begin{equation}
\vert\Psi\rangle \simeq \vert\Psi_{AB}\rangle +
\vert\Psi_{AB\gamma}\rangle + \vert\Psi_{AB\gamma\gamma}\rangle +
\vert\Psi_{AB\gamma\gamma\gamma}\rangle +\dots ,
\label{ABandphotons}
\end{equation}
where each component, involving an increasing number of additional
photons (or other possible additional particles), has a finite
non-vanishing norm that cannot be made arbitrarily small.
Typically, $\vert\vert\Psi_{AB}\vert\vert^2\simeq 1-o(\alpha)$,
$\vert\vert\Psi_{AB\gamma}\vert\vert^2\simeq o(\alpha)$,
$\vert\vert\Psi_{AB\gamma\gamma}\vert\vert^2\simeq o(\alpha^2)$,
although there may be additional suppression factors e.g.˜ due to
the ``phase space" limitations for the additional particles
\cite{pureprp}.

All the previous approaches to the EPR problem have assumed that
it was possible to produce with an arbitrary accuracy the state
$\vert\Psi_{AB}\rangle$, whose spin part is given by Eq.˜
(\ref{entanglement}). Now we see that this is merely a component
of the state vector given by Eq.˜ (\ref{ABandphotons}). The other
components do not imply energy, momentum and angular momentum
conservation for the two-particles A and B, since the additional
photons that they involve can carry these conserved quantities.
Therefore, {\it the measurement on A does not allow for a certain
prediction of the value of the considered conserved quantity} (be
it energy, momentum or angular momentum) {\it of B (B is not put
in an eigenstate of the observable that has been measured on A).}
For instance, $S_z$ {\it will not be given a ``physical reality"
on B after it is measured on the distant particle A.} Moreover,
the detection of particle A does not necessarily correspond to
particle B appearing in the opposite direction. After the
measurement, the amplitude for the additional undetected photons
fills the whole space, eventually overlapping with A and B;
therefore there is no theoretical possibility of defining two
determined spatially separated subsystems as required by the EPR
argument. Strictly speaking, it can be said that A and B
themselves are spatially separated only {\it after} measuring on
both particles, since the measurement of A in $O_A$ and global
momentum conservation are not sufficient to force B to appear in a
given direction (it is even possible that A and B are caught by
the same detector!). Therefore, the elements of physical reality
corresponding to the observables on B can be obtained only after
the {\it local} detection of B by $O_B$ \cite{pureprp}.

As a result, {\it QFT satisfies the EPR condition of completeness
without the need for additional hidden variables and without
violating the Einstein Locality assertion (2) in the EPR theorem,}
since the QFT state vector is not the two-particle state that was
understood in assertion (1). In other words, {\it the EPR paradox
is removed in an interpretation-independent way.} As far as I
know, {\it this is the first way out of the EPR theorem ever
found!} In particular, {\it the EPR+Bell proof of nonlocality is
also removed} \cite{locaqft}.

As I have discussed in the introduction, up to now the possible
solution to the EPR problem was a statistical interpretation based
on hidden variables. That option was discarded by Bell's theorem.
But now we see that QFT, without introducing QFT, does not imply
any EPR paradox or violation of Einstein Locality as formulated
above. Does this mean that it is a local theory? The answer
depends on the interpretation of the measurement process. If we do
not introduce hidden variables and do not go beyond QFT, we are
left with two possibilities:

I) The state vector applies to the single system. Now, since the
state vector of a free system evolves deterministically in Quantum
Mechanics and in QFT, the joint state of the measuring apparatus
(including all the ``environment" which interact with it) and the
object system after the measurement has to be determined by that
before the measurement. In particular, the position of the
pointer, i.e.˜ the result of the measurement, has to be the
deterministic consequence of the initial conditions, and the only
possible source of indeterminism is the unavoidable statistical
ignorance of the state of the environmental variables. Since the
QFT laws are local, the world would be intrinsically deterministic
and local. This possibility is logically consistent, but it seems
to be unable to provide a satisfactory solution to the
``measurement problem".

The ``orthodox" interpretations \cite{Laloe,Peres,dEspagnat}
introduce the collapse postulate in order to reconcile the
assumption that the state vector applies to the single system with
the fact that the measurement gives sharp results. The measurement
process is then considered a magical process, different from the
physical interactions that are well described by QFT. I think that
this ad hoc assumption is very unnatural in QFT; it also
contradicts the experience in particle physics detectors, that
shows that the measurement process is determined by the same
strong and electroweak interactions that are described by QFT.
Moreover, the collapse of the state vector has to simultaneously
involve all the space and explicitly violates locality and Lorentz
symmetry. This is unacceptable in a relativistic world, as
discussed above. Paradoxically, in spite of all these
contradictions and absurdities, orthodox interpretations are
currently the most common choice. Up to now, this may have been
justified precisely by the pressure of both EPR and Bell's
theorems taken together.

II) The remaining possibility for interpreting QFT without
introducing new physics is to assume that the state vector does
not describe the single system (which I will also call event), but
only describes an ensemble ${\cal E}$ of identically prepared
copies of the system (more precisely, statistical operators should
be used, since photons uncertainty prevents the preparation of a
pure state \cite{locaqft}). This minimal statistical
interpretation is clearly more economical than orthodox ones that
use the unnecessary assumption that the state vector also
describes the single systems. As a matter of fact, as a
consequence of the photons uncertainty, {\it QFT does not make any
prediction on the single event} (i.e.˜ on the single copy of the
system), {with the exception of the charges and masses of the
particles that will possibly appear} \cite{locaqft}. However, {\it
QFT predicts probability rates and cross sections that can be
compared with the frequencies of the results for the repetition of
an experiment on a statistical ensemble of equally-prepared copies
of the considered system.} This can be considered a hint in favor
of a statistical interpretation \cite{locaqft}.

Most importantly, such an interpretation naturally removes the
measurement problem \cite{Ballentine70,BJ,locaqft}, since the
state vector after the measurement of a magnitude ${\cal A}$ {\it
is} a linear combination involving the different eigenstates of
the observed physical quantity, as obtained with the linear QFT
laws. After the measurement, the experimenter (or the measuring
device) selects only the events that have given a particular
result, say $\alpha_1$, and this corresponds to considering a
subensemble ${\cal E}_1$ of the initial statistical ensemble
${\cal E}$. After this selection of the events, the state vector
is that which describes the new ensemble ${\cal E}_1$, which is
the same vector as in the usual collapse postulate (however, since
we are not associating a state to the individual system we do not
need a nonlinear evolution during the measurement process). This
also implies that {QFT with the statistical interpretation is a
fully local theory.} In fact, causality and locality are satisfied
by the QFT Green functions (and then by the scattering matrix)
\cite{WeinbookI}, and {\it there is no possible source of
nonlocality.} In particular, as I have shown in Refs.˜
\cite{pureprp,locaqft}, the prediction for the EPR correlations in
QFT is close to that calculated with the previous approach based
on Eq.˜ (\ref{entanglement}), and can be expressed in terms of
Lorentz-invariant Feynman amplitudes \cite{locaqft} (depending on
spin/helicities and momenta that transform under Lorentz
transformations as shown in Ref.˜ \cite{WeinbookI}). Bell's
inequalities are still violated, but this only implies that the
measurement of a spin component on A gives additional information
about the measurement of another spin component on B, with respect
to the information that can be contained in the state preparation
\cite{BaJa87,locaqft}. Ballantine and Jarrett \cite{BaJa87} have
proved that this fact should not be interpreted as a sign of
nonlocality, since it only implies that the quantum theory is less
predictive than a classical-looking theory.

Note that in Ref.˜ \cite{Ballentine70} the EPR theorem led to the
conclusion that the quantum theory had to be completed: the single
system had to have precise values of anticommuting observables
like momenta and positions, which played the role of hidden
variables. As I have commented in the introduction, this was
precisely the kind of solution that Einstein tended to favor,
however (at least in its radical version) it is ruled out by
Bell's theorem. Here, after removing EPR theorem, we do not need
to introduce hidden variables. Ironically, we have come to a
similar conclusion as EPR (the state vector does not describe the
single system), although now we do not need to provide a more
complete description! This fact may have important philosophical
consequences.

It has been claimed that in statistical interpretations which do
not introduce hidden variables ``all the systems that have the
same wave function are identical, since nothing differentiates
them" \cite{dEspagnat}. Such a criticism would be correct if we
did attribute the wave function to the single system. In a purely
probabilistic theory, the only conclusion that could be reached
with this kind of consideration would be that the minimal
statistical interpretation does not explain {\it why} the
measurements give sharp, different values, although it allows for
this. On the other hand, the unique known kind of interpretations
that would give a real ``explanation" would be those based on
hidden variables, which have problems with Bell's theorem (an
exception is Bohm's theory, which I will not consider here since
it has not been implemented to an interpretation of QFT). May be
that we will never solve this mystery, which might reach the
limits of science, although I hope that some progress will be
provided with a possible Theory of Everything. For the moment, QFT
with its statistical interpretation is not an {\it explanation,}
but at least it is a consistent {\it description}.

Therefore, the minimal statistical interpretation II) is the only
viable interpretation of QFT that does not rely on new physics.
The resulting theory is completely local. We have no guarantee
that a better understanding of the quantum correlations will ever
be found, however it can be hoped that this will eventually be
achieved in a possible ultimate Theory of Everything. In any case,
I think that locality and Lorentz symmetry will be preserved at
least as good approximations, and will not be so badly violated as
to allow whatever instantaneous effect between very distant
objects. In fact, it has been shown \cite{WeinbookI} that QFT is
the only reasonable form that a relativistic quantum theory should
have, and its enormous success in predicting the results of all
the present experiments suggests that it will be the good ``low
energy" approximation (corresponding precisely to the long
distance behavior) of any Theory of Everything.

Finally, I would like to mention the possible importance of these
results for Quantum Information Theory. First, the applications
that were based on the EPR paradox should not be interpreted as
signs of any kind of nonlocality. Secondly, the EPR correlations
that are usually assumed to be maximal (at least in principle) are
slightly reduced by photons uncertainty. This reduction is
particularly small if only the coincident events are selected by
local measurements on both particles A and B, thus reducing the
phase space available for additional photons. Since in Quantum
Cryptography applications this selection is usually performed, the
main consequence of photons uncertainty will be the fact that the
transmitted keys will {\it unavoidably} have some random errors
(some 1 will be read 0), with some small probability that can be
evaluated experimentally for the given settings. Such errors
should not be confused with a possible eavesdropper, which can be
ensured by using statistical tests such as that on Bell's
inequalities (which will be essentially unaffected by the photons
uncertainty). A remaining problem is that the random errors in the
final key cannot be avoided (Alice and Bob will not share exactly
the same key), but since they are rare, most of them can be
corrected simply by an orthographic corrector after ``translating"
the document (numbers should be written as full words, since a 3
instead of an 8 will not be corrected, but a eigft instead of an
eight will be caught); then, one should cross one's fingers. On
the other hand, teleportation should be reinterpreted as a
statistical process, while the non-cloning theorem can be
generalized to the case that the copied state is destroyed.

I thank H.˜ Michinel for very useful and stimulating discussions,
R.˜ Porto, U.˜ Trittmann and R.˜ Garc\'\i a Fern\'andez for
comments, and R.˜ Ramanathan for help. I am also very grateful to
Sergei Kilim and to the organizers of this interesting conference
for inviting me to deliver this talk.

\bibliography{puseprp}

\end{document}